\documentclass{elsart}
\newcommand{\be}{\begin{equation}}
\newcommand{\ee}{\end{equation}}
\newcommand{\beq}{\begin{eqnarray}}
\newcommand{\eeq}{\end{eqnarray}}
\newcommand{\mbf}{\mathbf}
\newcommand{\ds}{\displaystyle}

\begin{document}
\begin{frontmatter}
\title{Frequency decomposition of conditional Granger causality and application to multivariate
neural field potential data}

\author{Yonghong Chen}

\address{Department of Biomedical Engineering, University of Florida, Gainesville, FL 32611, USA}
\ead{ychen@bme.ufl.edu}

\author{Steven~L. Bressler}

\address{Center for Complex Systems and Brain Sciences, Florida Atlantic University, Boca
Raton, FL 33431, USA}
\ead{bressler@fau.edu}

\author{Mingzhou Ding}
\address{Department of Biomedical Engineering, University of Florida, Gainesville, FL 32611, USA}
\ead{mding@bme.ufl.edu}
\newpage
\begin{abstract}
It is often useful in multivariate time series analysis to
determine statistical causal relations between different time
series. Granger causality is a fundamental measure for this
purpose. Yet the traditional pairwise approach to Granger
causality analysis may not clearly distinguish between direct
causal influences from one time series to another and indirect
ones acting through a third time series. In order to differentiate
direct from indirect Granger causality, a conditional Granger
causality measure in the frequency domain is derived based on a
partition matrix technique. Simulations and an application to
neural field potential time series are demonstrated to validate
the method.
\end{abstract}

\begin{keyword}

Granger causality; Conditional Granger causality; Multiple time
series; Frequency domain; Multivariate autoregressive (MVAR) model;
Autoregressive moving average (ARMA) process; Partition matrix.
\end{keyword}
\end{frontmatter}

\newpage

\section{Introduction}
The concept of causality introduced by Wiener (Wiener, 1956) and
formulated by Granger (Granger, 1969) has played a considerable
role in investigating the relations among stationary time series.
The original definition of Granger (1969), which is well named as
Granger causality, refers to the improvement in predictability of
a series that derives from incorporation of the past of a second
series, above the predictability based solely on the past of the
first series. This definition only involves the relation between
two time series.  As pointed out by Granger (Granger, 1969; 1980),
if a third series is taken into account, a spurious or indirect
causality due to the third series may be detected.  Then he
defined a prima facie cause (Granger, 1980): $Y$ is said to be a
prima facie cause of $X$ if the observations of $Y$ up to time $t$
$(Y(\tau): \tau\leq t)$ help one predict $X(t+1)$ when the
corresponding observations of $X$ and $Z$ are available
$(X(\tau),Z(\tau): \tau\leq t)$. We refer to this idea as
conditional Granger causality since it gives a measure of
causality between two time series, $X$ and $Y$, conditional on a
third, $Z$. Evaluation of this conditional Granger causality in
the time domain is fairly straightforward through comparison of
two predictions of $X(t+1)$, one when $(X(\tau),Z(\tau): \tau\leq
t)$ are given, the other when $(X(\tau),Y(\tau),Z(\tau): \tau\leq
t)$ are given. However, evaluating causality by frequency
decomposition may allow more meaningful interpretations in cases
where oscillations are involved.

After giving clear measurements of linear dependence and feedback
between two blocks of time series (Geweke, 1982), Geweke also
presented a measure of conditional linear dependence and feedback
(Geweke, 1984).  Both a time domain measure, consistent with that
of Granger, and its frequency decomposition were given. Although
Hosoya presented some improvements on Geweke's methods (both
bivariate (Hosoya, 1991) and conditional versions (Hosoya, 2001)),
they have not been widely accepted because his time domain
implementation departs from Granger's original idea, and its
physical interpretation is less clear.

We point out that Geweke's use of the term "feedback" is
equivalent to "causality" in the present discussion. In applying
Geweke's frequency-domain conditional Granger causality measure to
neural field potential data, we have found that negative values,
which have no meaning in terms of causality, may occur at some
frequencies. This finding casts doubt on the applicability of
Geweke's method for neural time series analysis. We believe that
the negative values result from the lack of identity of estimates
of the same spectrum when different autoregressive (AR) models are
used. This non-identity of different estimates of the same
spectrum is a general practical problem in numerical analysis that
causes errors in Geweke's implementation because it requires the
estimates to be identical. In this paper, we employ a partition
matrix method to overcome this problem. Comparison of the results
from our procedure with Geweke's original procedure, clearly shows
the validity of the current procedure. In the following sections:
we first provide an introduction to Granger causality; then
present an overview of Geweke's procedure on conditional
causality, pointing out the importance of obtaining a correct
measure; and then derive our procedure. Finally, results of
simulations and application to neural field potential time series
data are provided.

\section{Background}
Consider a multiple stationary time series of dimension $n$, $\mbf
W=\{\mbf w_t\}$.  The series has the following moving average
representation with use of the lag operator $L$: \be \mbf w_t=\mbf
A(L) \mbox{\boldmath $\varepsilon_t$},\ee where $E(\mbox{\boldmath
$\varepsilon_t$})=\mbf 0, \mathrm{var}(\mbox{\boldmath
$\varepsilon_t$})=\mbf \Sigma$ and $\mbf A_0=\mbf I_n$, the $n
\times n$ identity matrix. Assume there exists the autoregressive
representation: \be \label{ar1} \mbf B(L)\mbf w_t=\mbox{\boldmath
$\varepsilon_t$},\ee where $\mbf B_0=\mbf I_n$.

Suppose that $\mbf w_t$ has been decomposed into two vectors $\mbf
x_t$ and $\mbf y_t$ with $k$ and $l$ dimensions respectively: $\mbf
w_t=(\mbf x_t',\mbf y'_t)'$, where the prime denotes matrix
transposition. Denote $\mbf W_{t-1}$ as the subspace generated by
$\{\mbf w_s;s\leq t-1\}$. Define $\mbf \Sigma_1=\mathrm{var}(\mbf
x_t|\mbf X_{t-1}), \mbf \Sigma_2=\mathrm{var}(\mbf x_t|\mbf
X_{t-1},\mbf Y_{t-1}), \mbf T_1=\mathrm{var}(\mbf y_t|\mbf Y_{t-1}),
\mbf T_2=\mathrm{var}(\mbf y_t|\mbf X_{t-1},\mbf Y_{t-1})$ and $\mbf
\Upsilon =\mathrm{var}(\mbf w_t|\mbf W_{t-1})$, where the
conditional variance is taken to be the variance of the residual
about the linear projection which accounts for the prediction. The
measures of linear causality from $\mbf Y$ to $\mbf X$, linear
causality from $\mbf X$ to $\mbf Y$, instantaneous linear causality
and linear dependence were respectively defined to be (Geweke,
1982):

\parbox{10cm}{\beq
& \ \ & F_{\mbf Y\rightarrow \mbf X} = \ln(|\mbf \Sigma_1|/|\mbf \Sigma_2|),\nonumber \\
& \ \ & F_{\mbf X\rightarrow \mbf Y} =  \ln(|\mbf T_1|/|\mbf T_2|),\nonumber \\
& \ \ & F_{\mbf X \cdot \mbf Y} = \ln(|\mbf \Sigma_2|\cdot|\mbf
T_2|/|\mbf
\Upsilon|),\nonumber \\
& \ \ & F_{\mbf X , \mbf Y} = \ln(|\mbf \Sigma_1|\cdot|\mbf
T_1|/|\mbf \Upsilon|) = F_{\mbf Y\rightarrow \mbf X}+F_{\mbf
X\rightarrow \mbf Y}+F_{\mbf X \cdot \mbf Y}. \nonumber \eeq}
\hfill
\parbox{1cm}{\beq  \eeq}

The measures of directional linear causality may be decomposed by
frequency. Let the autoregressive representation for $\mbf X$ and
$\mbf Y$ be:
\be\label{ar2}\left(%
\begin{array}{cc}
  \mbf B_{11}(L)\ \ & \mbf B_{12}(L) \\
  \mbf B_{21}(L)\ \ & \mbf B_{22}(L) \\
\end{array}%
\right)\left(%
\begin{array}{c}
  \mbf x_t \\
  \mbf y_t \\
\end{array}%
\right)=\left(%
\begin{array}{c}
  \mbox{\boldmath $\varepsilon_{1t}$} \\
  \mbox{\boldmath $\varepsilon_{2t}$} \\
\end{array}%
\right),\ee with $\mbf B_{11}(0)=\mbf I_k, \mbf B_{22}(0)=\mbf
I_l, \mbf B_{12}(0)=\mbf 0,\mbf B_{21}(0)=\mbf 0,
\mathrm{var}(\mbox{\boldmath $\varepsilon_{1t}$})=\mbf \Sigma_2,
\mathrm{var}(\mbox{\boldmath $\varepsilon_{2t}$})=\mbf T_2$.
Eq.(\ref{ar2}) is actually a partition form of Eq.(\ref{ar1}). Let
$\mbf C=\mathrm{cov}(\mbox{\boldmath
$\varepsilon_{1t}$},\mbox{\boldmath $\varepsilon_{2t}$})$. Then
pre-multiplying a transformation matrix \be \label{p} \mbf P=\left(%
\begin{array}{cc}
  \mbf I_k & \mbf 0 \\
  -\mbf C'\mbf \Sigma_2^{-1} \ \ & \mbf I_l \\
\end{array}%
\right)\ee to both sides of Eq.(\ref{ar2}), we have the following
normalized form:\be\label{ar33}\left(%
\begin{array}{cc}
  \tilde{\mbf B}_{11}(L)\ \ & \tilde{\mbf B}_{12}(L) \\
  \tilde{\mbf B}_{21}(L)\ \ & \tilde{\mbf B}_{22}(L) \\
\end{array}%
\right)\left(%
\begin{array}{c}
  \mbf x_t \\
  \mbf y_t \\
\end{array}%
\right)=\left(%
\begin{array}{c}
  \mbox{\boldmath $\varepsilon_{1t}$} \\
  \mbox{\boldmath $\tilde{\varepsilon}_{2t}$} \\
\end{array}%
\right),\ee where $\mbox{\boldmath $\varepsilon_{1t}$}$ and
$\mbox{\boldmath $\tilde{\varepsilon}_{2t}$}$ are uncorrelated,
$\mathrm{var}(\mbox{\boldmath $\tilde{\varepsilon}_{2t}$})=\mbf
T_3=\mbf T_2-\mbf C'\mbf \Sigma_2^{-1} \mbf C$, and $\tilde{\mbf
B}_{12}(0)=\mbf 0$ but $\tilde{\mbf B}_{21}(0)\neq\mbf 0$ in
general. Then Eq.(\ref{ar33}) implies the following spectral
decomposition of the spectral density of $X$:\be \mbf
{S_x}(\lambda)=\tilde{\mbf H}_{11}(\lambda)\mbf \Sigma_2 \tilde{\mbf
H}_{11}^*(\lambda)+\tilde{\mbf H}_{12}(\lambda)\mbf T_3 \tilde{\mbf
H}_{12}^*(\lambda),\ee where $\tilde{\mbf H}(\lambda)$ is the
transfer matrix of the normalized autoregressive expression in
Eq.(\ref{ar33}).  It is obvious that the spectral density of $\mbf
X$ is decomposed into an intrinsic part and a causal part, so the
measure of linear causality was suggested as (Geweke, 1982): \be
f_{\mbf Y\rightarrow \mbf X}(\lambda)=\ln\frac{|\mbf
{S_x}(\lambda)|}{|\tilde{\mbf H}_{11}(\lambda)\mbf \Sigma_2
\tilde{\mbf H}_{11}^*(\lambda)|}.\ee There is also a convergence
relation between the measures in the time and frequency domains:\be
\frac{1}{2\pi}\int_{\ \ -\pi}^{\ \
 \ \ \pi}f_{\mbf Y\rightarrow \mbf X}(\lambda)d \lambda \leq F_{\mbf Y\rightarrow \mbf X}.\ee

In the above, a Granger causality measure between two (or two
blocks of) time series was given. Before Geweke presented his
logarithm version, Pierce (Pierce, 1979) had introdued a $R^2$
measure which simply takes the ratio of the variances of two
prediction errors. The value of Pierce's $R^2$ measure is within
$[0,1]$ which is more convenient for comparison with correlation
coefficients. However, Geweke's logarithm version has better
statistical properties. There has also been  a measure based on
autoregressive moving average(ARMA) models (Boudjellaba et al.,
1992).

Granger causality analysis has been employed in a number of
studies of neural data (Bernasconi and K$\ddot{o}$nig, 1999;
Bernasconi et al., 2000; Kaminski et al., 2001; Hesse et al.,
2003; Harrison et al., 2003; Brovelli et al., 2004; Roebroeck et
al., 2005). The procedure described above has potential
applications in these types of study. For those cases where more
than two scalar/block time series recordings are available, the
procedure may be performed to identify further patterns of neural
interaction after a more traditional pairwise analysis. We now
consider two simple simulations to illustrate situations in which
conditional Granger causality analysis is important.

\subsection{Example 1: The case of differentially delayed driving}
We first consider a simple system consisting of three variables,
each representing an AR process:\be \label{Exar3}
\begin{array}{l}
x(t) = \xi(t) \\
y(t) = x(t-1)+\eta(t)\\
z(t) = \mu z(t-1)+x(t-2)+\epsilon(t).
\end{array} \ee
where $|\mu|<1$ is a parameter, and $\xi(t), \eta(t), \epsilon(t)$
are independent white noise processes with zero mean and variances
$\sigma_1^2,\sigma_2^2,\sigma_3^2$, respectively. The system
configuration is illustrated in Fig. 1(a) where $x$ drives $y$
after the delay of one time unit and $x$ drives $z$ after the
delay of two time units. We note that the time unit here is
arbitrary and has no physical meaning. To be consistent with the
data presented later we assume that the sample rate is 200 Hz. In
other words each time unit is 5 millisecond.

We performed a simulation of this system, with $\mu=0.5,
\sigma_1=1, \sigma_2=0.2,$ and $\sigma_3=0.3$, to generate a data
set of 500 realizations, each 100 points long. Then, assuming no
knowledge of Eq.(\ref{Exar3}), we fit multivariate autoregressive
(MVAR) models (Ding et al., 2000) to the generated data set for
each pairwise combination of variables $x$, $y$, and $z$, and
calculated the frequency-domain Granger causality for each pair in
each direction, as shown in Fig. 1(b). In the top two rows of this
figure, we see non-zero Granger causality values across the
spectra of ${\mbf x\rightarrow \mbf y}$ and ${\mbf x\rightarrow
\mbf z}$ and zero values across the spectra of ${\mbf y\rightarrow
\mbf x}$ and ${\mbf z\rightarrow \mbf x}$. These results are
indicative of the true unidirectional causal driving of $y$ and
$z$ by $x$. However, we also see results in the third row of Fig.
1(b) which appear to indicate unidirectional causal driving of $z$
by $y$. In fact, we know from the system configuration that this
apparent driving is due to the common influence of $x$ on both $y$
and $z$ but with different time delays. This mistaken
identification of an indirect influence as being a direct one
suggests the need for the conditional Granger causality measure.

\subsection{Example 2:  The case of sequential driving}
Next we consider another simple system, again consisting of three
AR processes: \be \label{3ar2}
\begin{array}{l}
x(t) = \xi(t) \\
y(t) = x(t-1)+\eta(t)\\
z(t) = \mu z(t-1)+y(t-1)+\epsilon(t).
\end{array} \ee
This system configuration consists of sequential driving from $x$ to
$y$, then from $y$ to $z$ as illustrated in Fig. 2(a). The same
numbers of realizations and data points were generated for the same
parameter values, and MVAR models were again fit to the data. The
results of Granger causality analysis in Fig. 2(b) show an apparent
unidirectional causal driving of $z$ by $x$ that is in fact due to
the indirect influence through $y$. Again, the mistaken
identification of an indirect influence as being direct suggests the
need for the conditional Granger causality measure.

Note that, although the systems in the above two examples are very
different, the results of pairwise Granger causality analysis seen
in Figs. 1(b) and 2(b) are essentially the same, indicating that
the analysis could not distinguish between the two systems. These
two examples, although simple, thus plainly demonstrate that the
pairwise measure of Granger causality by itself may be
insufficient to reveal true system relations. We now describe the
conditional Granger causality as a potentially useful tool for
disambiguating such situations.

\section{Geweke's Measure of Conditional Feedback Causality}
Now suppose that $\mbf w_t$ has been decomposed into three vectors
$\mbf x_t$, $\mbf y_t$ and $\mbf z_t$ with $k$, $l$ and $m$
dimensions, respectively: $\mbf w_t=(\mbf x_t',\mbf y'_t,\mbf
z'_t)'$.  The measure given by Geweke for the linear dependence of
$\mbf X$ on $\mbf Y$, conditional on $\mbf Z$, in the time domain
(geweke, 1984) is: \be \label{tmeasure} F_{\mbf Y\rightarrow \mbf
X|\mbf Z} = \ln {\frac{\mathrm{var}(\mbf x_t |\mbf X_{t-1},\mbf
Z_{t-1})}{\mathrm{var}(\mbf x_t |\mbf X_{t-1},\mbf Y_{t-1},\mbf
Z_{t-1})}}, \ee which is consistent with Granger's definition of a
prima facie cause (Granger, 1980).

Time series prediction is achieved by the fitting of MVAR models.
In order to implement Eq.(\ref{tmeasure}), two MVAR models are
involved. One is the following two-variable MVAR model:
\beq \label{xzdouble} \left(%
\begin{array}{cc}
  \mbf D_{11}(L) \ \ & \mbf D_{12}(L) \\
  \mbf D_{21}(L) \ \ & \mbf D_{22}(L) \\
\end{array}%
\right)\left(%
\begin{array}{c}
  \mbf x_t \\
  \mbf z_t \\
\end{array}%
\right) =\left(%
\begin{array}{c}
  \mbf \Theta_t \\
  \mbf \Psi_t \\
\end{array}%
\right), \eeq with the normalization $\mbf D_{11}(0)=\mbf I, \mbf
D_{22}(0)=\mbf I, \mbf D_{12}(0)=\mbf 0$, and $\mathrm{cov}(\mbf
\Theta_t,\mbf \Psi_t)=\mbf 0$ imposed in order to yield the
frequency decomposition of the conditional dependence. The
normalization can be achieved by using a transformation matrix
like Eq.(\ref{p}).

The other MVAR model used for deriving the frequency decomposition
of the conditional dependence is the following three-variable MVAR
model:

\beq \label{xyz} \left(%
\begin{array}{ccc}
 \mbf B_{11}(L) \ \ & \mbf B_{12}(L) \ \ & \mbf B_{13}(L) \\
 \mbf B_{21}(L) \ \ & \mbf B_{22}(L) \ \ & \mbf B_{23}(L) \\
 \mbf B_{31}(L) \ \ & \mbf B_{32}(L) \ \ & \mbf B_{33}(L) \\
\end{array}%
\right)\left(%
\begin{array}{c}
 \mbf x_t \\
 \mbf y_t \\
 \mbf z_t \\
\end{array}%
\right) =\left(%
\begin{array}{c}
 \mbox{\boldmath $\varepsilon_{xt}$} \\
 \mbox{\boldmath $\varepsilon_{yt}$} \\
 \mbox{\boldmath $\varepsilon_{zt}$} \\
\end{array}%
\right), \eeq with normalization imposed too.  The explicit formula
of the transformation matrix to normalize the MVAR model of three
time series is given in the Appendix.

Based on the relations of different variances, Geweke derived the
following important relation of the conditional causality in the
time domain (Geweke, 1984): \be F_{\mbf Y\rightarrow \mbf X|\mbf
Z}=F_{\mbf {Y\Psi}\rightarrow \mbf \Theta}. \ee The same
relationship is satisfied in the frequency domain (Geweke, 1984):\be
\label{relation} \textsl{f}_{\mbf Y\rightarrow \mbf X|\mbf
Z}(\lambda)=\textsl{f}_{\mbf {Y\Psi}\rightarrow \mbf
\Theta}(\lambda). \ee In order to get $\textsl{f}_{\mbf
{Y\Psi}\rightarrow \mbf \Theta}(\lambda)$, we need to decompose the
variance of $\mbf \Theta$ into the frequency domain. To do so, we
write Eq.(\ref{xzdouble}) and Eq.(\ref{xyz}) in the frequency
domain:
\beq \label{xzf} \left(%
\begin{array}{c}
  \mbf X(\lambda) \\
  \mbf Z(\lambda) \\
\end{array}%
\right) =\left(%
\begin{array}{cc}
  \mbf G_{xx}(\lambda) \ \ & \mbf G_{xz}(\lambda) \\
  \mbf G_{zx}(\lambda) \ \ & \mbf G_{zz}(\lambda) \\
\end{array}%
\right)\left(%
\begin{array}{c}
  \mbf \Theta(\lambda) \\
  \mbf \Psi(\lambda) \\
\end{array}%
\right), \eeq

\beq \label{xyzf} \left(%
\begin{array}{c}
  \mbf X(\lambda) \\
  \mbf Y(\lambda) \\
  \mbf Z(\lambda) \\
\end{array}%
\right) =\left(%
\begin{array}{ccc}
  \mbf H_{xx}(\lambda) \ \ & \mbf H_{xy}(\lambda) \ \ & \mbf H_{xz}(\lambda) \\
  \mbf H_{yx}(\lambda) \ \ & \mbf H_{yy}(\lambda) \ \ & \mbf H_{yz}(\lambda) \\
  \mbf H_{zx}(\lambda) \ \ & \mbf H_{zy}(\lambda) \ \ & \mbf H_{zz}(\lambda) \\
\end{array}%
\right)\left(%
\begin{array}{c}
  \mbf E_x(\lambda) \\
  \mbf E_y(\lambda) \\
  \mbf E_z(\lambda) \\
\end{array}%
\right). \eeq

If the spectra of $\mbf X(\lambda)$ and $\mbf Z(\lambda)$ from
Eq.(\ref{xzf}) remain identical to the spectra from
Eq.(\ref{xyzf}), then we can substitute Eq.(\ref{xzf}) into
Eq.(\ref{xyzf}) to get the following equations:
\beq \label{fdc} \left(%
\begin{array}{c}
  \mbf \Theta(\lambda) \\
  \mbf Y(\lambda) \\
  \mbf \Psi(\lambda) \\
\end{array}%
\right) &=& \left(%
\begin{array}{ccc}
  \mbf G_{xx}(\lambda) \ \ & \mbf 0 \ \ & \mbf G_{xz}(\lambda) \\
  \mbf 0 \ \ & \mbf I_l \ \ & \mbf 0 \\
  \mbf G_{zx}(\lambda) \ \ & \mbf 0 \ \ & \mbf G_{zz}(\lambda) \\
\end{array}%
\right)^{-1}\left(%
\begin{array}{ccc}
  \mbf H_{xx}(\lambda) \ \ & \mbf H_{xy}(\lambda) \ \ & \mbf H_{xz}(\lambda) \\
  \mbf H_{yx}(\lambda) \ \ & \mbf H_{yy}(\lambda) \ \ & \mbf H_{yz}(\lambda) \\
  \mbf H_{zx}(\lambda) \ \ & \mbf H_{zy}(\lambda) \ \ & \mbf H_{zz}(\lambda) \\
\end{array}%
\right)\left(%
\begin{array}{c}
  \mbf E_x(\lambda) \\
  \mbf E_y(\lambda) \\
  \mbf E_z(\lambda) \\
\end{array}%
\right)\nonumber \\
 &=& \left(%
\begin{array}{ccc}
  \mbf Q_{xx}(\lambda) \ \ & \mbf Q_{xy}(\lambda) \ \ & \mbf Q_{xz}(\lambda) \\
  \mbf Q_{yx}(\lambda) \ \ & \mbf Q_{yy}(\lambda) \ \ & \mbf Q_{yz}(\lambda) \\
  \mbf Q_{zx}(\lambda) \ \ & \mbf Q_{zy}(\lambda) \ \ & \mbf Q_{zz}(\lambda) \\
\end{array}%
\right)\left(%
\begin{array}{c}
  \mbf E_x(\lambda) \\
  \mbf E_y(\lambda) \\
  \mbf E_z(\lambda) \\
\end{array}%
\right), \eeq where
$\mbf{Q}(\lambda)=\mbf{G}^{-1}(\lambda)\mbf{H}(\lambda)$. From the
first equation of Eqs.(\ref{fdc}), the spectrum of $\mbf \Theta$
is decomposed into the following three obvious parts:\be \mbf S_{
\Theta}(\lambda)=\mbf Q_{xx}(\lambda)\mbf \Sigma_{xx}\mbf
Q_{xx}^*(\lambda)+\mbf Q_{xy}(\lambda)\mbf \Sigma_{yy}\mbf
Q_{xy}^*(\lambda)+\mbf Q_{xz}(\lambda)\mbf \Sigma_{zz}\mbf
Q_{xz}^*(\lambda).\ee Therefore the measure of causality from
$\mbf {Y\Psi}$ to $\mbf \Theta$ may be described as:\be
\textsl{f}_{\mbf {Y\Psi}\rightarrow \mbf \Theta}(\lambda)=\ln
\frac{\left|\mbf S_{\Theta}(\lambda)\right|}{\left|\mbf
Q_{xx}(\lambda)\mbf \Sigma_{xx}\mbf Q_{xx}^*(\lambda)\right|},\ee
where $\mbf S_{\Theta}(\lambda)$ is actually the variance of $\mbf
\Theta_t$, namely $\mbf \Sigma_{\Theta}$, since $\mbf \Theta_t$ is
white noise in Eq.(\ref{xzdouble}). Considering the relation of
Eq.(\ref{relation}), we could get the conditional causality as
\cite{geweke84}: \be \label{cm} \textsl{f}_{\mbf Y\rightarrow \mbf
X|\mbf Z}(\lambda)=\ln \frac{|\mbf \Sigma_{\Theta}|}{\left|\mbf
Q_{xx}(\lambda)\mbf \Sigma_{xx}\mbf Q_{xx}^*(\lambda)\right|},\ee

In the above derivations, the assumption that the spectra of $\mbf
X(\lambda)$ and $\mbf Z(\lambda)$ coming from Eq.(\ref{xzf}) and
from Eq.(\ref{xyzf}) are identical is actually very hard to
satisfy numerically due to practical estimation errors. As an
example of this problem, consider Fig. 6, where the dashed curves
result from performing Geweke's conditional casuality procedure.
Note that the negative values seen here have no interpretation in
terms of causality. (A detailed description of Fig. 6 is given in
a later section.) In the following section, we introduce the
partition matrix technique to overcome this problem.

\section{Partition Matrix Improvement}
For three blocks of time series $\mbf x_t$, $\mbf y_t$, $\mbf z_t$,
we can fit a three-variable MVAR model as in Eq.(\ref{xyz}) and we
can also derive its frequency domain expression as in
Eq.(\ref{xyzf}). From Eq.(\ref{xyzf}), writing an expression only
for $\mbf X(\lambda)$ and $\mbf Z(\lambda)$ (making
partitions) we have:\beq \label{xzfr} \left(%
\begin{array}{c}
  \mbf X(\lambda) \\
  \mbf Z(\lambda) \\
\end{array}%
\right) =\left(%
\begin{array}{cc}
  \mbf H_{xx}(\lambda) \ \ & \mbf H_{xz}(\lambda) \\
  \mbf H_{zx}(\lambda) \ \ & \mbf H_{zz}(\lambda) \\
\end{array}%
\right)\left(%
\begin{array}{c}
  \bar{\mbf E}_x(\lambda) \\
  \bar{\mbf E}_z(\lambda) \\
\end{array}%
\right), \eeq where $\bar{\mbf E}_x(\lambda)$ and $\bar{\mbf
E}_z(\lambda)$ have the
following moving average expression: \beq \label{re} \left(%
\begin{array}{c}
  \bar{\mbf E}_x(\lambda) \\
  \bar{\mbf E}_z(\lambda) \\
\end{array}%
\right) =\left(%
\begin{array}{c}
  \mbf E_x(\lambda) \\
  \mbf E_z(\lambda) \\
\end{array}%
\right)+\left(%
\begin{array}{cc}
  \mbf H_{xx}(\lambda) \ \ & \mbf H_{xz}(\lambda) \\
  \mbf H_{zx}(\lambda) \ \ & \mbf H_{zz}(\lambda) \\
\end{array}%
\right)^{-1}\left(%
\begin{array}{c}
  \mbf H_{xy}(\lambda) \\
  \mbf H_{zy}(\lambda) \\
\end{array}%
\right)\mbf E_y(\lambda). \eeq

We realize that Eq.(\ref{xzfr}) is actually a summation of multiple
ARMA processes, and that the summation of several ARMA processes is
still an ARMA process (Granger and Morris, 1976; Harvey, 1993).
However, an unambiguous representation of the general multivariate
ARMA process for the summation is unknown, although a general
univariate ARMA model could be obtained through a specific procedure
(Maravall and Mathis, 1994). Alternatively, we adopt the following
procedure to evaluate the conditional Granger causality.

Letting \beq \left(%
\begin{array}{c}
  \bar{\mbf H}_{xy}(\lambda) \\
  \bar{\mbf H}_{zy}(\lambda) \\
\end{array}%
\right)=\left(%
\begin{array}{cc}
  \mbf H_{xx}(\lambda) \ \ & \mbf H_{xz}(\lambda) \\
  \mbf H_{zx}(\lambda) \ \ & \mbf H_{zz}(\lambda) \\
\end{array}%
\right)^{-1}\left(%
\begin{array}{c}
  \mbf H_{xy}(\lambda) \\
  \mbf H_{zy}(\lambda) \\
\end{array}%
\right) \nonumber, \eeq we get the covariance matrix of the noise
terms given in
Eq.(\ref{re}):\beq \label{covm} \bar{\mbf \Sigma}(\lambda) = \left(%
\begin{array}{cc}
 \mbf  \Sigma_{xx} \ \ & \mbf \Sigma_{xz} \\
 \mbf  \Sigma_{zx} \ \ & \mbf \Sigma_{zz} \\
\end{array}%
\right) & + & \left(%
\begin{array}{c}
  \bar{\mbf H}_{xy}(\lambda) \\
  \bar{\mbf H}_{zy}(\lambda) \\
\end{array}%
\right)\left(\mbf \Sigma_{xy} \ \ \mbf \Sigma_{zy}\right)
 + \left(%
\begin{array}{c}
  \mbf \Sigma_{xy} \\
  \mbf \Sigma_{zy} \\
\end{array}%
\right)\left(\bar{\mbf H}_{xy}^*(\lambda) \ \ \bar{\mbf
H}_{zy}^*(\lambda)\right)\nonumber \\ & + &
\mbf \Sigma_{yy} \left(%
\begin{array}{c}
  \bar{\mbf H}_{xy}(\lambda) \\
  \bar{\mbf H}_{zy}(\lambda) \\
\end{array}%
\right) \left(\bar{\mbf H}_{xy}^*(\lambda) \ \ \bar{\mbf
H}_{zy}^*(\lambda)\right).\eeq This covariance matrix is no longer
a real matrix, but it is a Hermite matrix, i.e. $\bar{\mbf
\Sigma}_{xz}(\lambda)=\bar{\mbf \Sigma}_{zx}^*(\lambda)$.
Therefore we can use the following transformation matrix to
normalize the bivariate
model of Eq.(\ref{xzfr}):\be \bar{\mbf P}=\left(%
\begin{array}{cc}
  \mbf I_k \ \ \ & \mbf 0 \\
  -\ds \frac{\bar{\mbf \Sigma}_{xz}(\lambda)} {\bar{\mbf \Sigma}_{xx}} \ \ \ & \mbf I_m \\
\end{array}%
\right). \ee Therefore, in correspondence with the normalized form
in Eq.(\ref{xzf}), the transfer matrix $\mbf G(\lambda)$ is now:
\be \mbf
G(\lambda)=\left(%
\begin{array}{cc}
  \mbf H_{xx}(\lambda)\ \ & \mbf H_{xz}(\lambda) \\
  \mbf H_{zx}(\lambda)\ \ & \mbf H_{zz}(\lambda) \\
\end{array}%
\right)\left(%
\begin{array}{cc}
  \mbf I_k \ \ \ & \mbf 0 \\
  -\ds \frac{\bar{\mbf \Sigma}_{xz}(\lambda)} {\bar{\mbf \Sigma}_{xx}} \ \ \ & \mbf I_m \\
\end{array}%
\right)^{-1} \ee Taking the expansion form of this $\mbf
G(\lambda)$ matrix to get matrix $\mbf Q(\lambda)=\mbf
G^{-1}(\lambda)\mbf H(\lambda)$, and considering $\mbf
\Sigma_\Theta=\bar{\mbf \Sigma}_{xx}$, where $\bar{\mbf
\Sigma}_{xx}$ comes from Eq.(\ref{covm}), we can still use
Eq.(\ref{cm}) to get the conditional causality.

\section{Applications to Simulated and Neural Field Potential Data}
\subsection{Application to Simulated Data}

We performed conditional Granger causality analysis on the delay
driving and sequential driving systems presented above in Section
$2$. For the delay driving case (Section 2.1 and Fig. 1), the
Granger causality spectrum from $y$ to $z$, conditional on $x$, is
presented in Fig. 3. It is obvious from Fig. 3 that the
conditional Granger causality measure eliminated the indirect
causal influence of $y$ on $z$ which appeared in Fig. 1(b). For
the sequential driving case (Section 2.2 and Fig. 2), the Granger
causality from $x$ to $z$, conditional on $y$, is also presented
in Fig. 3. Clearly, the indirect causal influence from $x$ to $z$,
which was indicated in Fig. 2(b), was also eliminated by use of
the conditional Granger causality.

In both cases, we have seen that conditional Granger causality
analysis eliminated indirect causal influences that inadvertently
resulted from application of the pairwise Granger causality
measure. Knowing the system equations in these examples allowed us
to verify that the conditional Granger causality measure yielded a
truer depiction of the system relations. We now consider how the
conditional Granger causality measure may provide the same benefit
in the analysis of real neural data.

\subsection{Application to Neural Field Potential Data}

Field potential data were recorded from two macaque monkeys using
transcortical bipolar electrodes at 15 distributed sites in
multiple cortical areas of one hemisphere (right hemisphere in
monkey GE and left hemisphere in monkey LU) while the monkeys
performed a GO/NO-GO visual pattern discrimination task (Bressler
et al., 1993). The presence of oscillatory field potential
activity in the beta (14-30 Hz) frequency range was recently
reported in the sensorimotor cortex of these monkeys during the
prestimulus period (Brovelli et al., 2004). In that study, Granger
causality analysis was performed for all pairwise combinations of
sensorimotor cortical recording sites. In both monkeys,
significant Granger causal influences were discovered from primary
somatosensory cortex to both primary motor cortex and inferior
posterior parietal cortex, with the latter area also exerting
Granger causal influences on primary motor cortex.

In monkey GE, the possibility existed that the causal influence
from the primary somatosensory ($Soma$) site to one of the
inferior posterior parietal sites (in area 7a) was actually
mediated by another inferior posterior parietal site (in area 7b)
(Fig. 4(a)). We therefore used conditional Granger causality
analysis to test the hypothesis that the ${ Soma\rightarrow 7a}$
influence was mediated by the $7b$ site. In Fig. 4(b) is presented
the pairwise Granger causality spectrum from the $Soma$ site to
the $7a$ site (${ Soma\rightarrow 7a}$, light solid curve),
showing significant causal influence in the beta frequency range.
Superimposed in Fig. 4(b) is the conditional Granger causality
spectrum for the same pair, but with the $7b$ site taken into
account (${Soma \rightarrow 7a|7b}$, dark solid curve). The
corresponding 99\% significance thresholds are also presented
(light and dark dashed lines which overlap each other). These
significance thresholds were determined using a permutation
procedure that involved creating 500 permutations of the field
potential data set by random rearrangement of the trial order.
Since the test was performed separately for each frequency, a
correction was necessary for the multiple comparisons over the
whole range of frequencies. The Bonferroni correction could not be
employed because these multiple comparisons were not independent.
An alternative strategy was employed following (Blair and
Karniski, 1993). The Granger causality spectrum was computed for
each permutation, and then the maximum causality value over the
frequency range was identified. After 500 permutation steps, a
distribution of maximum causality values was created. Choosing a
p-value at $p=0.01$ for this distribution gave the thresholds
shown in Fig. 4(b), (c) and Fig. 5(b) in dashed lines.

We see from Fig. 4(b) that the conditional Granger causality is
greatly reduced in the beta frequency range and no longer
significant, meaning that the causal influence from the $Soma$ site
to the $7a$ site is most likely an indirect effect mediated by the
$7b$ site. This conclusion is consistent with the known neuroanatomy
of the sensorimotor cortex (Felleman and Essen, 1991) in which area
7a is connected with area 7b, but not directly with the primary
somatosensory cortex.

From Fig. 4(a) we see that the possibility also existed that the
causal influence from the $Soma$ site to the primary motor ($Mot$)
site in monkey GE was mediated by the $7b$ site. To test this
possibility, the Granger causality spectrum from $Soma$ to $Mot$
(${Soma \rightarrow Mot}$, light solid curve in Fig. 4(c)) was
compared with the conditional Granger causality spectrum with $7b$
taken into account (${Soma \rightarrow Mot|7b}$, dark solid curve in
Fig. 4(c)). In contrast to Fig. 4(b), we see that the beta-frequency
conditional Granger causality in Fig. 4(c) is only partially
reduced, and remains well above the 99\% significance level. In Fig.
5(a), we see that the same possibility existed in monkey LU of the
$Soma$ to $Mot$ causal influence being mediated by $7b$. However,
just as in Fig. 4(c), we see in Fig. 5(b) that the beta-frequency
conditional Granger causality for monkey LU is only partially
reduced, and remains well above the 99\% significance level.

The results from both monkeys thus indicate that the Granger
causal influence from the primary somatosensory cortex to the
primary motor cortex was not simply an indirect effect mediated by
area 7b. However, we further found that area 7b did play a role in
mediating the $Soma$ to $Mot$ causal influence in both monkeys.
This was determined by comparing the means of bootstrap resampled
distributions of the peak beta Granger causality values from the
spectra of ${ Soma\rightarrow Mot}$ and ${ Soma\rightarrow
Mot|7b}$ by Student's t-test. The significant reduction of
beta-frequency Granger causality when area 7b is taken into
account (t = 17.2 for GE; t = 18.2 for LU, p $<<<$ 0.001 for
both), indicates that the influence from the primary somatosensory
to primary motor area was partially mediated by area 7b. Such an
influence is consistent with the known neuroanatomy (Felleman and
Essen, 1991), which shows direct connections between area 7b and
both primary motor and primary somatosensory areas.

As a final demonstration of the value of using the partition
matrix method outlined in this paper to compute conditional
Granger causality, we present in Fig. 6 a direct comparison of our
improved procedure (solid) with Geweke's original procedure
(dashed) for the ${ Soma\rightarrow Mot|7b}$ spectra of monkey GE
(Fig. 6(a)) and monkey LU (Fig. 6(b)). From much previous
experience working with this field potential data (Brovelli et
al., 2004), we know that spectra from these cortical areas
typically have a single peak in the beta frequency range. Geweke's
original method is clearly seen to be deficient in these examples
not only by the multiple peaks and valleys across the spectra, but
also by the negative values, which have no physical
interpretation. We thus are confident that the partition matrix
technique is a potentially valuable tool to be used in the
investigation of conditional Granger causality relations between
neural signals.

\section*{Acknowledgments}

The work was supported by US NIMH grants MH64204, MH070498 and
MH71620, and NSF grant 0090717.

\section*{Appendix: Transformation matrix to normalize a model of three time series}
Since the MVAR model, such as in Eq.(\ref{xyz}), is usually not
normalized, the noise terms could be correlated with each other.
Let us assume that the covariance matrix is given by: \beq \nonumber \mathbf{\Sigma} = \left(%
\begin{array}{ccc}
  \mbf \Sigma_{xx} \ \ &  \mbf \Sigma_{xy} \ \ &  \mbf \Sigma_{xz} \\
   \mbf \Sigma_{yx} \ \ &  \mbf \Sigma_{yy} \ \ &  \mbf \Sigma_{yz} \\
   \mbf \Sigma_{zx} \ \ &  \mbf \Sigma_{zy} \ \ &  \mbf \Sigma_{zz} \\
\end{array}%
\right).  \eeq In order to make the first noise term independent,
we
could use the following transform: \beq \nonumber \mathbf{P}_1=\left(%
\begin{array}{ccc}
   \mbf I_k \ \ &  \mbf 0 \ \ &  \mbf 0 \\
  - \mbf \Sigma_{yx} \mbf \Sigma_{xx}^{-1} \ \ &  \mbf I_l \ \ \ &  \mbf 0 \\
  - \mbf \Sigma_{zx} \mbf \Sigma_{xx}^{-1} \ \ & 0 \ \ \ &  \mbf I_m \\
\end{array}%
\right).\eeq Then the covariance matrix for the transformed noise
terms is:

\beq \nonumber \left(%
\begin{array}{ccc}
   \mbf \Sigma_{xx} \ \ &  \mbf 0 \ \ &  \mbf 0 \\
   \mbf 0 \ \ &  \mbf \Sigma_{yy}- \mbf \Sigma_{yx} \mbf \Sigma_{xx}^{-1} \mbf \Sigma_{xy} \ \ &  \mbf \Sigma_{yz}-\mbf \Sigma_{yx} \mbf \Sigma_{xx}^{-1} \mbf \Sigma_{xz} \\
   \mbf 0 \ \ & \mbf \Sigma_{zy}-\mbf \Sigma_{zx} \mbf \Sigma_{xx}^{-1} \mbf \Sigma_{xy} \ \ & \mbf \Sigma_{zz}-\mbf \Sigma_{zx} \mbf \Sigma_{xx}^{-1} \mbf \Sigma_{xz} \\
\end{array}%
\right).  \eeq

Again, to make the second and third noise terms independent, the
following transformation may be made:
 \beq \nonumber \mathbf{P}_2=\left(%
\begin{array}{ccc}
  \mbf I_k \ \ & \mbf 0 \ \ & \mbf 0 \\
  \mbf 0 \ \ & \mbf I_l \ \ \ & \mbf 0 \\
  \mbf 0 \ \ \ & -(\mbf \Sigma_{zy}-\mbf \Sigma_{zx}\mbf \Sigma_{xx}^{-1}\mbf \Sigma_{xy})(\mbf \Sigma_{yy}-\mbf \Sigma_{yx}\mbf \Sigma_{xx}^{-1}\mbf \Sigma_{xy})^{-1} \ \ &  \mbf I_m \\
\end{array}%
\right).\eeq Therefore the whole transformation matrix needed to
make all three noise terms independent is:  \beq \nonumber
\mathbf{P}= \mathbf{P}_2 \cdot \mathbf{P}_1\eeq

\newpage

\section*{Figure Caption}

\begin{description}
\item{\bf Figure 1:} (a) Schematic illustration of a simple delay
driving system; (b) Granger causality spectra by pairwise
analysis.

\item{\bf Figure 2:} (a) Schematic illustration of a simple
sequential driving system; (b) Granger causality spectra by
pairwise analysis.

\item{\bf Figure 3:} Conditional Granger causality spectra from
$y$ to $z$, conditional on $x$, in the delay driving example (Fig.
1), and from $x$ to $z$, conditional on $y$, in the sequential
driving example (Fig. 2), showing that the indirect effects
indicated in Figs. 1(b) and 2(b) have been eliminated.

\item{\bf Figure 4:} Granger causality analysis of field potential
data from monkey GE. (a) Schematic illustration of significant
Granger causal influences in the beta frequency band among sites
in sensorimotor cortex. (b) The reduction of the conditional
Granger causality spectrum (dark solid line) below the
significance threshold indicates an indirect influence of the
primary somatosensory site on the 7a site. (c) The fact that the
conditional Granger causality spectrum (dark solid line) does not
fall below the significance threshold shows that there is a direct
influence of the primary somatosensory site on the primary motor
site. That the conditional spectrum is significantly below the
pairwise spectrum indicates that there is an additional indirect
effect mediated by 7b.

\item{\bf Figure 5:} Granger causality analysis of field potential
data from monkey LU.  (a) Schematic illustration of significant
Granger causal influences in the beta frequency band among sites
in sensorimotor cortex. (b) As in Fig. 4(c), the conditional
Granger causality spectrum (dark solid line) does not fall below
the significance threshold, indicating that there is a direct
influence of the primary somatosensory site on the primary motor
site. Again, the conditional spectrum is significantly below the
pairwise spectrum, showing that there is an additional indirect
effect mediated by 7b.

\item{\bf Figure 6:} Comparison of Geweke's original method for
computing conditional Granger causality spectra (dashed) with our
partition matrix procedure (solid). (a) ${ Soma\rightarrow
Mot|7b}$ in monkey GE, corresponding to Fig. 4(c); (b) ${
Soma\rightarrow Mot|7b}$ in monkey LU, corresponding to Fig. 5(b).
Note that for both monkeys, Geweke's original method suffers from
having multiple peaks and valleys across the spectra (believed to
be artifactual), and also from having negative values, which have
no physical interpretation.

\end{description}

\end{document}